\documentclass[aps,twocolumn,showpacs,floatfix]{revtex4}
\usepackage{amsmath}
\usepackage{amsfonts}
\usepackage{amssymb}
\usepackage{graphicx}
\begin{document}
\title{Quantum Entanglement Initiated Super Raman Scattering}
\author{G. S. Agarwal}
\affiliation{Department of Physics,Oklahoma State University, Stillwater,
Oklahoma 74078, USA}

\date{\today}
\begin{abstract}
It has now been possible to prepare chain of ions in an entangled state and thus question arises --- how the optical properties of a chain of entangled ions differ from say a chain of independent particles. We investigate nonlinear optical processes in such chains. We explicitly demonstrate the possibility of entanglement produced super Raman scattering. Our results in contrast to Dicke's work on superradiance are applicable to stimulated processes and are thus free from the standard complications of multimode quantum electrodynamics. Our results suggest the possibility of similar enhancement factors in other nonlinear processes like four wave mixing.
\end{abstract}
\maketitle

Dicke in 1954 \cite{Dicke} predicted that if a collection of two level atoms is prepared in a collective state where half the atoms are in the excited state and half the atoms are in the ground state then the spontaneous emission from such an initial state is proportional to the square of the number of atoms \cite{Allen}. Dicke introduced collective description of a system of $N$ atoms. The collective system can be considered in the state $|\frac{N}{2},M \rangle$ where as usual $M=-\frac{N}{2},\cdots,+\frac{N}{2}$ in steps of unity and $\frac{N}{2}$ is the net spin of the collection of atoms. The state $|\frac{N}{2},0\rangle$ exhibits superradiance. The origin of superradiance  is difficult to see as in the state $|\frac{N}{2},0\rangle$ there is no macroscopic dipole moment whereas such a dipole is required for radiation rate that is proportional to $N^{2}$. It turns out that the superradiant state $|\frac{N}{2},0\rangle$ has very interesting quantum correlations. Denoting the spin operator associated with each two level atom as $\vec{s}_{i}$, then the correlation between the $i^{th}$ atom and $j^{th}$ atom can be shown to be \cite{Agarwal}
\begin{equation}\label{1}
\langle s_{i}^{+}s_{j}^{-}\rangle-\langle s_{i}^{+}\rangle\langle s_{j}^{-}\rangle\cong1/4.
\end{equation}

The radiation rate in Dicke's work is proportional to $\sum_{i,j}\langle s_{i}^{+}s_{j}^{-}\rangle$ which would be proportional to $N^{2}$ as all the atomic correlations are of the order of unity. Thus it is clear that the Dicke superradiance arises from quantum correlations in the state that he introduced. Somehow the fact that in the Dicke state there are strong quantum correlations between the individual atoms has not been appreciated much until recently. There is considerable revival of interest in superradiance. Scully and coworkers have been examining the case of single photon superradiance \cite{Scully1,Scully2,Scully3,Scully4}. There are also studies \cite{Chaudhury} on the behavior of large spins (value 3) in electromagnetic fields. It also turns out that the Dicke states have very strong quantum entanglement. The entanglement character is easy to see for the case of two two level atoms with the two states labeled as $|e \rangle,|g \rangle$. In this case the state $|1,0\rangle$ that Dicke considered would be $|1,0\rangle = \frac{1}{\sqrt{2}}[|e_{1},g_{2}\rangle+|g_{1},e_{2}\rangle]$. This state, also known as the Bell state or the EPR state, is clearly entangled. For three atoms one of the Dicke states is $|3/2,1/2 \rangle$ which in current language would be W-state $\frac{1}{\sqrt{3}}[|e_{1},g_{2},g_{3}\rangle+|g_{1},e_{2},g_{3}\rangle+|g_{1},g_{2},e_{3}\rangle]$. The W state is also known to be fully entangled and this state is important in considerations of single photon superradiance \cite{Scully1,Scully2,Scully3,Scully4}. Thus a number of recent works \cite{Thiel,Prevedel,Wieczorek,Hume,Lemr,Linington} have also shown how the Dicke states for a small number of atoms can be prepared in laboratory.  The question that we address in this letter is--are there processes such as nonlinear optical processes other than spontaneous emission where systems prepared in Dicke like states can lead to enhancements in the efficiency of the process. As a new paradigm in the study of nonlinear optical processes we consider the well known process of Raman scattering and demonstrate how its efficiency can be enhanced depending on the quantum entanglement in the initial state in which atoms are prepared. We initiate our demonstration of super Raman scattering using the preparation of atoms in a W state. We then present the most general argument for super Raman scattering. We give enhancement factors in a number of special cases. Some of our results can be tested using the W state of chains of trapped ions \cite{Roos,Chou} while other results depend on the progress towards entanglement of qutrits. It would be clear from our discussion that a whole body of nonlinear processes can be enhanced by using entangled states of the medium.

\begin{figure}[htp]
 \scalebox{0.7}{\includegraphics{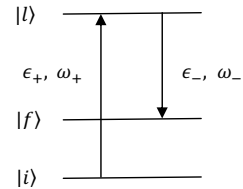}}
\caption{Stimulated Raman scattering. $|i\rangle$, $|l\rangle$, and $|f\rangle$ represent the initial state, the intermediate state, and the final state, respectively. A circular polarized photon with polarization $\epsilon_{+}$ and frequency $\omega_{+}$ is absorbed, and a circular polarized photon with polarization $\epsilon_{-}$ and frequency $\omega_{-}$ is emitted. All the states $|i\rangle$, $|l\rangle$, and $|f\rangle$ are the entangled states of the $N$ atom system.}
\end{figure}
Consider the second order process of stimulated Raman scattering as shown in the Fig. 1. Usual method of the calculation of the transition probability consists of the examination of the transitions in a single atom. The final transition probability is obtained by multiplying the result by the number of atoms as one assumes that the initial state is a factorized state. We proceed differently as the states that we deal with are entangled states. The transition probability for the Raman process can be calculated by second order perturbation theory \cite{Loudon}, where we replace in the standard expression various states by the entangled states
\begin{eqnarray}\label{2}
\Gamma_{R}=\frac{2\pi}{\hbar}\left|\sum_{l}\frac{\langle i|H_{1}|l\rangle \langle l|H_{1}|f\rangle}{E_{i}-E_{l}+\hbar\omega_{+}}\right|^{2}\delta(E_{i}+\hbar\omega_{+}-E_{f}-\hbar\omega_{-}),\nonumber\\
\end{eqnarray}
in which $E_{l}$ is the energy of the entangled $l^{th}$ state and $H_{1}$ is the interaction between the atomic system and the radiation fields. The $H_{1}$ has the form --$\vec{d}\cdot\vec{E}$;
\begin{equation}\label{3}
\vec{E}(\vec{r})=E_{+}\hat{\epsilon}_{+}e^{i\vec{k}_{l}\cdot\vec{r}}+E_{-}\hat{\epsilon}_{-}e^{i\vec{k}_{s}\cdot\vec{r}}+c.c.,
\end{equation}
and $\vec{d}$ is the dipole moment operator for the $N$ atom system:
\begin{equation}\label{4}
\vec{d}=\sum_{\alpha,i,j}(\vec{d})_{ij}|i\rangle_{\alpha\alpha}\langle j|+c.c.
\end{equation}
The sum in (\ref{4}) is over all the states and the states of the $\alpha^{th}$ atom are denoted by $|i\rangle_{\alpha}$. Clearly the dependence of $\Gamma_{R}$ on the number of atoms is determined by the matrix elements in (\ref{2}). Before we derive the general result we like to illustrate our key idea with a simple example. This would show us how the enhancement can arise. Consider for illustration first a system of three atoms which are prepared in an entangled W-state
\begin{eqnarray}\label{5}
|i\rangle=\frac{1}{\sqrt{3}}(|g_{-},g_{-},g_{+}\rangle+|g_{-},g_{+},g_{-}\rangle+|g_{+},g_{-},g_{-}\rangle),
\end{eqnarray}
in which any two atoms are in the state $|g_{-}\rangle$. We are specifically choosing Zeeman states so that we make use of the light polarization to select specific transitions. Let us consider the absorption of a photon with polarization $\epsilon_{+}$ and scattering of a photon with polarization $\epsilon_{-}$. Clearly an atom in the state $|g_{-}\rangle$ would be transferred to the state $|g_{+}\rangle$ via this scattering event. The final state would be one in which two atoms are in the state $|g_{+}\rangle$. Thus we expect the final state to be a W-state of the type
\begin{eqnarray}\label{6}
|f\rangle=\frac{1}{\sqrt{3}}(|g_{+},g_{-},g_{+}\rangle+|g_{+},g_{+},g_{-}\rangle+|g_{-},g_{+},g_{+}\rangle).
\end{eqnarray}
It should be borne in mind that now we are investigating Raman process which starts from an entangled state and ends at an entangled state. Since the efficiency of the Raman process is determined by the nature of all the states involved in the transition, we next examine the intermediate state in the process, which clearly will be
\begin{eqnarray}\label{7}
& &\frac{1}{\sqrt{3}}(|e_{o},g_{-},g_{+}\rangle e^{i\vec{k}_{l}\cdot\vec{R}_{1}}+|g_{-},e_{o},g_{+}\rangle e^{i\vec{k}_{l}\cdot\vec{R}_{2}}\nonumber\\& &+|e_{o},g_{+},g_{-}\rangle e^{i\vec{k}_{l}\cdot\vec{R}_{1}}+|g_{-},g_{+},e_{o}\rangle e^{i\vec{k}_{l}\cdot\vec{R}_{3}}\nonumber\\& &+|g_{+},e_{o},g_{-}\rangle e^{i\vec{k}_{l}\cdot\vec{R}_{2}}+|g_{+},g_{-},e_{o}\rangle e^{i\vec{k}_{l}\cdot\vec{R}_{3}}).
\end{eqnarray}
Here $\vec{R}_{i}$ is the position of the $i^{th}$ atom and $\vec{k}_{l}$ is the wavevector of the field with polarization $\epsilon_{+}$.
 The final state obtained from (\ref{7}) by the scattering of $\epsilon_{-}$ would be
\begin{eqnarray}\label{8}
& &\frac{1}{\sqrt{3}}(|g_{+},g_{-},g_{+}\rangle e^{i\vec{k}_{l}\cdot\vec{R}_{1}-i\vec{k}_{s}\cdot\vec{R}_{1}}\nonumber\\& &+|g_{-},g_{+},g_{+}\rangle e^{i\vec{k}_{l}\cdot\vec{R}_{2}-i\vec{k}_{s}\cdot\vec{R}_{2}}\nonumber\\
& &+|g_{+},g_{+},g_{-}\rangle e^{i\vec{k}_{l}\cdot\vec{R}_{1}-i\vec{k}_{s}\cdot\vec{R}_{1}}\nonumber\\
& &+|g_{-},g_{+},g_{+}\rangle e^{i\vec{k}_{l}\cdot\vec{R}_{3}-i\vec{k}_{s}\cdot\vec{R}_{3}}\nonumber\\& &+|g_{+},g_{+},g_{-}\rangle e^{i\vec{k}_{l}\cdot\vec{R}_{2}-i\vec{k}_{s}\cdot\vec{R}_{2}}\nonumber\\
& &+|g_{+},g_{-},g_{+}\rangle e^{i\vec{k}_{l}\cdot\vec{R}_{3}-i\vec{k}_{s}\cdot\vec{R}_{3}})\nonumber\\&=&2\cdot\frac{1}{\sqrt{3}}(|g_{+},g_{-},g_{+}\rangle\left(\frac{e^{i\varphi_{1}}+e^{i\varphi_{3}}}{2}\right)\nonumber\\& &+|g_{-},g_{+},g_{+}\rangle\left(\frac{e^{i\varphi_{2}}+e^{i\varphi_{3}}}{2}\right)\nonumber\\& &+|g_{+},g_{+},g_{-}\rangle\left(\frac{e^{i\varphi_{1}}+e^{i\varphi_{2}}}{2}\right)),
\end{eqnarray}
in which $\varphi_{\alpha}=(\vec{k}_{l}-\vec{k}_{s})\cdot\vec{R}_{\alpha}$, and $\alpha=1,2,3$. If the scattering geometry is chosen so that the incident and scattered fields lie in a plane perpendicular to the plane of atoms, then the final state is the entangled state.

Note the presence of the numerical factor 2 in Eq. (\ref{8}). This would lead to a factor 4 in the transition probability. We have thus shown that the transition probability for 3 atoms is four times than that for a single atom if the atoms are prepared initially in the entangled state (\ref{5}). For the standard case where atoms are prepared in the factorized state $|g_{-}\cdots g_{-}\rangle$ the transition probability would be three times than that for a single atom.  Clearly the use of the entangled W state has produced an enhanced Raman transition the enhancement factor being 4/3 for three atom system. For $N$ atoms prepared initially in a W state this argument leads to a factor $2(N-1)$ instead of the factor $N$ $(N\geq2)$ in the transition probability. Thus the usage of the entangled W state enhances the transition probability for the Raman process by a factor $\mathcal{E}=2-2/N$. This prediction can possibly be tested using the W state for 8 trapped ions as in the work of Innsbruck group \cite{Roos}.

We next consider the scattering cross section for the second order process following an argument similar to Dicke's using collective operators and collective states. Let us consider essentially a three level scheme for the second order process with the initial state $|i\rangle$, the final state $|f\rangle$, and the intermediate state $|l\rangle$. Let us introduce the collective states $|n_{i},n_{l},n_{f}\rangle$, where $n_{l}$ atoms are in the state $|l\rangle$. Note that $n_{i}+n_{l}+n_{f}=N$ \cite{Puri}. Let us introduce the collective operators $S_{ij}$, defined in terms of the operators for individual atoms as
\begin{eqnarray}\label{9}
S_{ij}=\sum_{\alpha}|i\rangle_{\alpha \alpha}\langle j|.
\end{eqnarray}
These collective operators satisfy the SU(3) algebra,
\begin{eqnarray}\label{10}
[S_{il},S_{fj}]=S_{ij}\delta_{lf}-S_{fl}\delta_{ij};
\end{eqnarray}
and can be represented in terms of the harmonic oscillators as $S_{il}=a_{i}^{\dag}a_{l}$, where $a_{i}'^{s}$ satisfy Bosonic commutation relations. From this representation it can be shown that these collective operators have the properties
\begin{eqnarray}\label{11}
S_{fi}|n_{i},n_{l},n_{f}\rangle=\sqrt{n_{i}(n_{f}+1)}|n_{i}-1,n_{l},n_{f}+1\rangle,\nonumber\\
S_{li}|n_{i},n_{l},n_{f}\rangle=\sqrt{n_{i}(n_{l}+1)}|n_{i}-1,n_{l}+1,n_{f}\rangle,
\end{eqnarray}
etc. For second order Raman process, we have
\begin{eqnarray}\label{12}
|n_{i},n_{l},n_{f}\rangle\stackrel{\omega_{+}}\rightarrow |n_{i}-1,n_{l}+1,n_{f}\rangle\stackrel{\omega_{-}}\rightarrow |n_{i}-1,n_{l},n_{f}+1\rangle.\nonumber\\
\end{eqnarray}
Hence the numerical factor giving the enhancement over single atom would be
\begin{eqnarray}\label{13}
\mathcal{E}=n_{i}(n_{f}+1)(n_{l}+1)^{2}/N.
\end{eqnarray}
Several important conclusions can be drawn from Eq. (\ref{13}). First of all if all the atoms are in the initial state ie $n_{i}=N$, $n_{f}=n_{l}=0$; then as expected $\mathcal{E}=1$. For an initial W state with $n_{i}=N-1$, $n_{l}=0$, $n_{f}=1$, $\mathcal{E}= 2-2/N$; a result derived in Sec II. Further if  $n_{l}=0$, $n_{i}=n_{f}=N/2$ then $\mathcal{E}= N(N+2)/(4N)\sim N$. The last result is reminiscent of the Dicke result on superradiance.

A new aspect of the enhancement for second order process is the appearance of the factor $(n_{l}+1)^{2}$. This suggests that much larger enhancement factors are possible by preparing the system initially in a collective state such that part of the population is present in the intermediate state $|l\rangle$. In particular if we consider Dicke like states involving the ground state and the intermediate state $|l\rangle$ ie $n_{i}=n_{l}=N/2$ then $\mathcal{E}$ is of the order of $N^{2}$. This exceptional enhancement comes from the fact that the entanglement properties of the state are used twice once in absorption and once in emission. Further note that if we start with a W state prepared such that $n_{i}=N-1, n_{l}=1, n_{f}=0$ then the enhancement factor (\ref{13}) becomes $\mathcal{E}=4(1-\frac{1}{N})$.

We conclude this paper by discussing further possibilities. We have shown how the entangled character of the states involved in the second order process leads to enhancement in the efficiency of the optical processes. Our results on W states basically use the entanglement among the qubits as the states involved in the transition are either $|i\rangle$, $|f\rangle$, or $|i\rangle$, $|l\rangle$. Such an entanglement among the qubits has been extensively studied \cite{Roos,Chou,Blatt} and can thus be utilized in testing some of our predictions. The more general case involves entanglement of qutrits which is beginning to be studied \cite{Inoue,Langford,Zeilinger,Lima,Bertlmann,Checinska}. It is clear from our results on second order Raman processes that other nonlinear optical processes in systems prepared in suitable entangled states would exhibit enhancement over that obtained in systems without entanglement. More generally our results point out to the need of developing the theory of nonlinear optical susceptibilities for systems prepared in entangled states as the conventional susceptibilities are derived assuming a model where the atoms interact with fields in an independent fashion \cite{Boyd}. Finally we note that it would also be interesting to examine efficiencies for nonlinear optical processes by replacing the classical fields by entangled fields \cite{Mukamel}.

The author is grateful to Mike Steiner, A. K. Rajagopal, M. Frey for discussions on the question of extending Dicke's work to other process, and to M. Scully for extensive discussions on single photon superradiance. The author thanks the hospitality of the Director, Indian Institute for Science Education and Research, Pune where this work was partly done.

\end{document}